\documentclass[11pt,a4paper]{article}
\pdfoutput=1
\usepackage{jcappub}
\usepackage{float}
\usepackage{graphicx}
\usepackage{epsfig}
\usepackage{epstopdf}
\usepackage{indentfirst}
\newcommand{\be}{\begin{equation}}
\newcommand{\ee}{\end{equation}}

\newcommand{\ba}{\begin{eqnarray}}
\newcommand{\ea}{\end{eqnarray}}

\begin{document}

\title{A de Sitter tachyon thick braneworld}
\author[a]{Gabriel Germ\'an}
\author[a,b]{Alfredo Herrera--Aguilar}
\author[a,b]{Dagoberto Malag\'on--Morej\'on}
\author[b]{Refugio Rigel Mora--Luna}
\author[c]{Rold\~ao da Rocha}
\affiliation[a]{Instituto de Ciencias F\'{\i}sicas, Universidad
Nacional Aut\'onoma de M\'exico.
Apdo. Postal 48-3, 62251, Cuernavaca, Morelos, M\'{e}xico.}
\affiliation[b]{Instituto de F\'{\i}sica y Matem\'{a}ticas, Universidad
Michoacana de San Nicol\'as de Hidalgo.
Edificio C--3, Ciudad Universitaria, C.P. 58040, Morelia,
Michoac\'{a}n, M\'{e}xico.}
\affiliation[c]{Centro de Matem\'atica, Computa\c c\~ao e Cogni\c c\~ao,
Universidade Federal do ABC\\
Rua Santa Ad\'elia, 166 09210-170, Santo Andr\'e, SP, Brazil}
\emailAdd{gabriel@fis.unam.mx}
\emailAdd{aha@fis.unam.mx}
\emailAdd{malagon@ifm.umich.mx}
\emailAdd{rigel@ifm.umich.mx}
\emailAdd{roldao.rocha@ufabc.edu.br}

\abstract{Among the multiple 5D thick braneworld models that have been proposed in the last years,
in order to address several open problems in modern physics, there is a specific one  involving a
tachyonic bulk scalar field. Delving into this framework, a thick braneworld with a cosmological
background induced on the brane is here investigated. The respective field equations --- derived
from the model with a warped 5D geometry --- are highly non-linear equations, admitting a
non-trivial solution for the warp factor and the tachyon scalar field as well, in a de Sitter 4D
cosmological background. Moreover, the non-linear tachyonic scalar field, that generates the brane
in complicity with warped gravity, has the form of a kink-like configuration. Notwithstanding, the
non-linear field equations restricting character does not allow one to easily find thick brane
solutions with a decaying warp factor which leads to the localization of 4D gravity and other
matter fields. We derive such a thick brane configuration altogether in this tachyon-gravity
setup. When analyzing the spectrum of gravity fluctuations in the transverse traceless sector, the
4D gravity is shown to be localized due to the presence of a {\it single} zero mode bound state,
separated by a continuum of massive Kaluza-Klein (KK) modes by a mass gap. It  contrasts with
previous results, where there is a KK massive bound excitation providing no clear physical
interpretation. The mass gap is determined by the scale of the metric parameter $H$. Finally, the
corrections to Newton's law in this model are computed and shown to decay exponentially. It is in
full compliance to corrections reported in previous results (up to a constant factor) within
similar braneworlds with induced 4D de Sitter metric, despite the fact that the warp factor and
the massive modes have a different form.}

\maketitle

\flushbottom


\section{Introduction}

Within  the framework  of  the braneworld models embedded in a spacetime with extra dimensions and
after the success of the thin brane models  --- where singularities are present at the position of
the branes --- in solving the mass hierarchy and 4D gravity localization problems \cite{gog,rs},
to find smooth braneworld solutions has cogently become a matter of interest (for an interesting
review see, e. g., \cite{thbrs} and references therein). In some models, such solutions are
obtained by introducing one or several scalar fields in the bulk. The large variety of scalar
fields that can be used to generate these models elicit different scenarios \cite{dewolfe, gremm,
csakietal, Giovaninni, KKS, fRwaveBW, ariasybarbosaetal, SR, Bazeiaetal, koleykar}. By following
this direction, and by using the freedom to choose a scalar field, several authors have been
evoking a tachyonic scalar field in the bulk \cite{Bazeiaetal, koleykar, senguptaetal2}. They are
furthermore concerned to address issues like the mass hierarchy problem, and localization of
gravity and matter fields, in both the thin and the thick branes models as well. In the original
Randall-Sundrum (RS) model, a Standard Model or TeV brane is introduced at a certain fixed
distance, say $r_c$, from the gravitational or Planck brane, in order to achieve the desired
warping, and hence, solve the hierarchy problem in a completely 5D geometrical way. However, this
resolution mechanism impels to a new fine-tuning on the probe brane position, and therefore, to
the need of stabilizing this brane separation. The stabilization of this brane separation is
achieved through the Goldberger-Wise mechanism, by associating to it a radion scalar field that
models the radius of the fifth dimension, when one ignores the brane back-reaction \cite{GW}.
This mechanism was further generalized to the case when one takes into account the brane
back-reaction in \cite{dewolfe}. Moreover, in \cite{senguptaetal1} the authors proposed a
tachyonic scalar field action for modeling and stabilizing the brane separation for the full
back-reacted system. They also obtained the desired warping from the Planck scale to the TeV one,
resolving the fine tuning problem of the Higgs mass in a stable braneworld scenario and
generalizing in a relevant way the RS model as well (see also \cite{senguptaetal2}). This
fact physically motivates the use of a tachyonic scalar field within the braneworld paradigm,
since the back-reaction of the radion field must be taken into account in a self--consistent
system.

On the another hand, the thick brane configuration constructed in \cite{palkar} possesses an
increasing warp factor, since most of the attempts to solve the highly non--linear field equations
leads to imaginary tachyon field configurations. This fact translates into delocalization of
gravity and other kinds of matter, like scalar and vector fields, while giving rise to
localization of fermions. Alternatively, the localization of fermionic fields on thick
branes was investigated in \cite{zld} with an auxiliary scalar field that couples to the fermionic 
mass term.

In this paper we propose a new thick brane tachyonic solution, with a decaying warp factor that
enables localization of 4D gravity as well as other matter fields. By analyzing the dynamics of
the metric perturbations, we realize that their spectrum contains a {\it single} bound state
corresponding to the 4D massless graviton of the model. Furthermore, it presents a continuum of KK
excitations separated by a mass gap. We compute in addition the corresponding correction to
Newton's law, by analyzing the influence of these massive modes on the gravitational potential
acting between two point massive particles, located along the center of the thick brane.

\section{The thick brane model and its solution}

The complete action for the tachyonic braneworld model is expressed
as
\begin{equation}
S = \int d^5 x \sqrt{-g} \left(\frac{1}{2\kappa_5^2} R - \Lambda_5\right) - \int d^{5}x \sqrt{-g}
V(T)\sqrt{1+g^{AB}\partial_{A} T\partial_{B} T} \label{accion}
\end{equation}
where the first term describes 5D gravity with a bulk cosmological constant $\Lambda_5,$ the
second is the action of the matter in the bulk, $\kappa_5$ is the 5D gravitational coupling
constant, and $A,B=0,1,2,3,5$. Hence, the tachyon field $T$ represents the matter in the 5D bulk
and $V(T)$ denotes its self--interaction potential \cite{bergshoeff,sen}.

The tachyon action part in Eq. (\ref{accion}) above was proposed in \cite{bergshoeff} within the
context of a tachyon field living on the world volume of a non--BPS brane and found applications
in braneworld cosmology \cite{mazumdar} and in string cosmology \cite{gibbons}. It was argued in
\cite{padmanabhan} that this form of the action can be used for any relativistic scalar field.
Moreover, using this action as a scalar tensor theory, solar system constraints were analyzed in
\cite{devi}.

The Einstein equations with a cosmological constant in five dimensions are given by
\begin{equation}
 G_{AB} = - \kappa_5^2 ~\Lambda_5 g_{AB} + \kappa_5^2 ~T_{AB}^{\it{bulk}}.
\label{einequ}
\end{equation}
For the background metric, the ansatz of a warped 5D line
element, with an induced 3--brane in a spatially flat cosmological
background, is used
\begin{equation}
ds^2 = e^{2f(\sigma)} \left[- d t^2 + a^2(t) \left(d x^2 + d y^2 + d
z^2 \right)  \right] + d \sigma^2, \label{ansatz}
\end{equation}
where $f(\sigma)$ is the warp factor and $a(t)$ is the scale factor
of the brane.

The matter field equation is obtained by variation of the action with respect to the tachyon. It
is expressed in the following form:
\begin{equation}
\Box T-\frac{\nabla_C\nabla_D T\,\, \nabla^{C} T\,\, \nabla^{D}
T}{1+(\nabla T)^2}= \frac{1}{V} \frac{\partial V(T) }{\partial T}.
\label{fieldequ1}
\end{equation}
By using the ansatz (\ref{ansatz}),
the Einstein tensor components read
\begin{eqnarray}
G_{00} &=& 3\, \frac{\dot a^2}{a^2} - 3\, e^{2 f}\left( f^{''} + 2 f^{'2} \right),
\label{eqeintach1}\nonumber \\
G_{\alpha\alpha} &=& - 2\, \ddot aa - \dot a^2 + 3\, a^2 e^{2 f} \left( f^{''} + 2 f^{'2} \right)
\label{eqeintach2},
\nonumber\\
G_{\sigma\sigma} &=& -3\, e^{-2 f} \left( \frac{\ddot a}{a} + \frac{\dot a^2}{a^2} \right) +
6f^{'2}, \label{eqeintach3}
\end{eqnarray}
where ${``\prime"}$ and ${``\cdotp"}$ are the derivative with respect to the extra dimension and
time, respectively, while $\alpha$ labels the spatial dimensions $x,$ $y$ and $z$. The
 stress energy tensor components do not depend explicitly on the time component, and read
\begin{equation}
T_{AB}^{\it{bulk}} =  \left[ - g_{AB} \, V(T) \sqrt{1 +
    (\nabla T)^2} + \frac {V(T)}{\sqrt{1+ (\nabla T)^2}} \,
\partial_{A} T \, \partial_{B} T \right].
\end{equation}
Since the non--diagonal components of the Einstein tensor vanish, consistency of Einstein
equations demands that non--diagonal components of the stress energy tensors should vanish
identically. This allows two possibilities: (a) the field $T$ should depend merely on time and not
on any of the spatial coordinates --- which is the case for a scalar field in an homogeneous and
isotropic background as in cosmology; (b) the field $T$ depends only on the coordinate
corresponding to the extra dimension.  This simply amounts to a consistent time independence of
the tachyon field, even if the background is time dependent, and shall be considered here, since we
are not regarding cosmology. Hence Eq.(\ref{fieldequ1}) reads
\begin{equation}
T^{''}+4f^{'}T^{'}(1+T^{'2})=(1+T^{'2})\frac{\partial_{T}V(T) }{V(T)}. \label{fieldequ}
\end{equation}
while the Einstein equations (\ref{einequ}) can be rewritten in a
straightforward way:
\begin{eqnarray}
f^{''} &=& - \kappa_5^2\frac{V(T)T^{'2}}{3\sqrt{1 + T^{'2}}} - e^{-2 f}\, \frac{\ddot a}{a},
\label{einsteinequ} \\
f^{'2} &=& - \kappa_5^2\frac{V(T)}{6\sqrt{1 + T^{'2}}} - ~\frac{\kappa_5^2\,\Lambda_5}{6} +
\frac{e^{-2 f}}{2} \left(\frac{\ddot a}{a}+\frac{\dot a^2}{a^2} \right). \label{restriccion}
\end{eqnarray}
Thereupon the case of a single brane configuration, in a 4D spatially flat cosmological
background given by equations (\ref{fieldequ}), (\ref{einsteinequ}), and (\ref{restriccion}), is evinced. The
consistency between the component ${\footnotesize  \alpha\alpha}$ of the Einstein equation with Eqs. (\ref{einsteinequ})
and (\ref{restriccion}) demands that the scale factor has to be $a(t)=ke^{H\,t}$, where $H$ and
$k$ are integration constants. Therefore, it corresponds to a de Sitter 4D cosmological background
defined by
\begin{equation}
a(t)=e^{H\,t}, \label{scalefactor}
\end{equation}
since the constant $k$ can be absorbed into a coordinate redefinition. This result is dictated
simply by the symmetry of the background. Thus, the action for the tachyonic scalar field is a
non--trivial 5D configuration that leads to a braneworld in which the induced metric on the brane
is described by $dS_4$ geometry.

Here it is convenient to go to conformal coordinates through $dw=e^{-f(y)}d\sigma$, leading to
the following metric
\begin{equation}
ds^2 = e^{2f(w)} \left[- d t^2 + a^2(t) \left( dx^2 + dy^2 + dz^{2}\right) + dw^2\right].
\label{ansatzconf}
\end{equation}
In the language of these coordinates the field equations (\ref{fieldequ}), (\ref{einsteinequ}) and
(\ref{restriccion}) can be written as follows
\begin{eqnarray}
T^{''}-f^{'}T^{'}+4f^{'}T^{'}(1+e^{-2 f}T^{'2})&=&(e^{2 f}+T^{'2})\frac{\partial_{T}V(T) }{V(T)}.
\label{Tz}\\
f^{''} - f^{'2} + H^2 &=& - \kappa_5^2\frac{V(T)T^{'2}}{3\sqrt{1 + e^{-2 f}T^{'2}}},
\label{Ee1z} \\
f^{'2} + \frac{\kappa_5^2\,\Lambda_5}{6}e^{2 f} - H^2 &=& - \kappa_5^2\frac{e^{2 f}V(T)}{6\sqrt{1
+ e^{-2 f}T^{'2}}}. \label{Ee2z}
\end{eqnarray}
where now the symbol ${}^{``\prime"}$ stands for derivatives with respect to $w$, and the form of the scale
factor corresponding to (\ref{scalefactor}) was taken into account.

According to the authors of \cite{senguptaetal1}, it is straightforward to obtain separate
equations for the scalar field $T$ and for the potential $V(T)$ from Eqs.(\ref{Ee1z}) and
(\ref{Ee2z}) by the following procedure: divide (\ref{Ee1z}) into (\ref{Ee2z}), isolate $T^{'2}$
and compute the square root of both sides of the equation; in order to get an explicit expression
for $T(w)$, one must be able to integrate the resulting expression. The self--interaction potential
$V(T(w))$ can be obtained in a parametric form by isolating this quantity from (\ref{Ee2z}) after 
replacing the corresponding expression for $T^{'2}$ in the radicand. If one is able to invert
$T(w)$,  it is possible hence to express the potential $V$ as a function of $T$.

Thus, it turns out that despite the high non--linearity of these field equations, the derivative
of the tachyonic scalar field $T^{'}$, as well as the arbitrary potential $V(T)$, can be expressed
in terms of the warp and scale factors of the metric (and their respective derivatives), after
some simple manipulations, leading to:
\begin{eqnarray}
T'&=&\pm
e^f\sqrt{\frac{f''-f'^2+H^2}{2\left(f'^2+\frac{\kappa_5^2\,\Lambda_5}{6}\,e^{2f}-H^2\right)}},
\label{Tprime}\\
V(T)&=& - \frac{3}{\kappa_5^2}\,e^{-2f}\,\sqrt{\frac{2\left(f''+
f'^2+\frac{\kappa_5^2\,\Lambda_5}{3}\,e^{2f}-H^2\right)} {\left(f'^2 +
\frac{\kappa_5^2\,\Lambda_5}{6}\,e^{2f}-H^2\right)}}
\left(f'^2+\frac{\kappa_5^2\,\Lambda_5}{6}\,e^{2f}-H^2\right). \label{V}
\end{eqnarray}
Therefore, by determining a desired behavior for the geometry, the dynamics of the tachyon field
is completely fixed, and vice--versa. However, the resulting solution must be real and have
physical sense. This restriction is nonetheless cogently demanding, since several warp factors
with ``convenient" behavior lead to a complex tachyonic field $T$ and/or self--interaction
potential $V(T)$.

It is straightforward to realize that the following warp factor
\begin{equation}
f(w)=-\frac{1}{2}\ln\left[\frac{\cosh\left[\,H\,(2w+c)\right]}{s}\right], \label{fw}
\end{equation}
where $H$, $c$ and $s>0$ are constants, is a solution of the Einstein and field equations, if the
tachyon scalar field adopts the form
\begin{equation}
T(w) = \pm\sqrt{\frac{-3}{2\,\kappa_5^2\,\Lambda_5}}\
\mbox{arctanh}\left[\frac{\sinh\left[\frac{H\,\left(2w+c\right)}{2}\right]}
{\sqrt{\cosh\left[\,H\,(2w+c)\right]}}\right], \label{Tw}
\end{equation}
and the tachyon potential  given by the following expression
\begin{eqnarray}
V(T) &=& - \Lambda_5\ \mbox{sech}\left(\sqrt{-\frac{2}{3}\kappa_5^2\,\Lambda_5}\ T\right)
\sqrt{6\ \mbox{sech}^2\left(\sqrt{-\frac{2}{3}\kappa_5^2\,\Lambda_5}\ T\right)-1}\nonumber \\
&=& - \frac{\Lambda_5}{\sqrt{2}}\ \sqrt{1+\mbox{sech}\left[\,H\,(2w+c)\right]} \sqrt{2+3\
\mbox{sech}\left[\,H\,(2w+c)\right]}. \label{VT}
\end{eqnarray}
In the last two equations we set
\begin{equation}
s=-\frac{6H^2}{\kappa_5^2\,\Lambda_5} \label{s}
\end{equation}
with a negative bulk cosmological constant $\Lambda_5<0$ for consistency.

\begin{figure}[htb]
\begin{center}
\includegraphics[width=7cm]{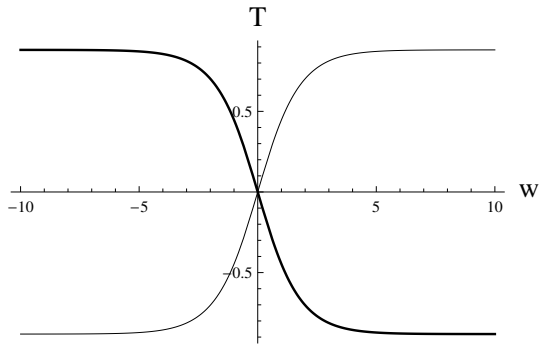}
\end{center}\vskip -5mm
\caption{The profile of the tachyonic scalar field $T$. The thin line represents the positive
branch of the field and the thick line is associated to the negative  branch of the tachyon.  Here
we have set $c=0,$ $H=1/2$, $2\kappa_{5}^2=1$ and $\Lambda_5=-3$ for simplicity.} \label{fig_T}
\end{figure}

\begin{figure}[htb]
\begin{center}
\includegraphics[width=7cm]{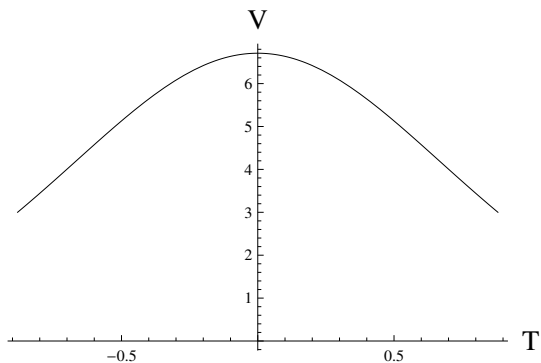}
\end{center}\vskip -5mm
\caption{The shape of the self--interaction potential of the tachyonic scalar field $V(T)$. We set
$\Lambda_5=-3$ and $2\kappa_{5}^2=1$ for simplicity. Due to the
bounded character of the tachyonic field, the potential
does not extends to infinity.} \label{fig_VT}
\end{figure}

Therefore, the set of equations (\ref{scalefactor}), (\ref{fw}), (\ref{Tw}), and (\ref{VT}) are
the necessary and sufficient ingredients that provide  a tachyonic thick brane solution. The
warp factor has a decaying and vanishing asymptotic behavior, whereas the tachyon scalar is real
and possesses a kink or antikink--like profile as shown in Fig. 1. In contrast with solutions
found in \cite{palkar}, here we manage to express the tachyon potential $V(T)$ in terms of the
tachyonic scalar field $T$. This potential has a maximum/minimum at the position of the brane, but
it is positive/negative definite as it can be seen from (\ref{VT}). In fact, since the tachyon
field is bounded, the potential remains real as well as bounded (see Fig. 2).

It is worth noticing that due to the relation (\ref{s}), we can compute the limit of this field
configuration when the bulk cosmological constant vanishes. In order to do this, we consider the
limit when $\Lambda_5 \to 0$ with the same rapidity as $H^2 \to 0$ does in such a way that $s$
remains finite. In this limit, the tachyon field becomes linear $T=\pm\frac{\sqrt{s}}{4}(2w+c)$,
the self--interaction potential vanishes $V(T)=0$, while the scale and warp factors become
constant that can be absorbed into the metric coordinates, leading to a flat 5D spacetime. This
situation is similar to that which arises in the RS model \cite{rs}, among others
\cite{dewolfe,gremm,csakietal,cuco}, when taking such a limit. For instance, in the RS solution
the warp factor becomes equal to unity, while the brane tensions vanish, leading to a flat
spacetime as well.

By computing the 5D curvature scalar for our solution (\ref{fw})
\begin{equation}
R=-\frac{14}{3}\kappa_5^2\,\Lambda_5\,\mbox{sech}\left[\,H\,(2w+c)\right] \label{R5}
\end{equation}
we see that this 5D invariant is positive definite and asymptotically vanishes, yielding an
asymptotically 5D Minkowski spacetime \cite{mannheim}. In the absence of matter, usually the
presence of a negative cosmological constant $\Lambda_5$ leads to an asymptotically $AdS_5$
spacetime. However, if we look at the equation (\ref{accion}) and/or (\ref{einequ}), we see that
both the cosmological constant $\Lambda_5$ and the self--interaction potential $V(T)$ contribute to the
overall effective cosmological constant of the 5D spacetime of our scalar tensor setup.

\section{Gravity localization and corrections to Newton's law}

In order to study the metric and field fluctuations of our system, both must be perturbed. Since
the relevant geometry of the four--dimensional background is of de Sitter type, the fluctuations
of the metric may be classified into tensorial, vector, and scalar modes, with respect to the
transformations associated to the symmetry group $dS_4$ \cite{spectroscopy}. It is of prominent
importance, since at first (linear) order these modes evolve independently. Hence, their dynamical
equations decouple, even when the perturbed Einstein and tachyon field equations had been highly
coupled non--linear equations. According to it, the dynamics of the tensorial metric fluctuations
is now studied,  which have the physical interpretation of the 5D braneworld graviton. These
tensorial metric fluctuations are gauge invariant and enable us forthwith to determine whether the
localization of 4D gravity, and hence the physics of our 4D world, is feasible or not within this
model.

To begin with, the Einstein equations are written in the form:
\begin{equation}
R_{A B}=\frac{2}{3}\kappa_5^2\,\Lambda_{5} g_{A B} +\kappa_{5}^{2} \hat{T}_{A B},
\label{ecualternativa}
\end{equation}
where the reduced energy--momentum tensor
\begin{equation}
\hat{T}_{A B}= \frac{2}{3}V\sqrt{1+(\nabla T)^2} g_{A B} - \frac{V}{\sqrt{1+(\nabla T)^2}}\left[
\frac{1}{3}{g_{A B}}(\nabla T)^2 -\nabla_{A} T \nabla_{B} T \right] \label{redTmn}
\end{equation} is employed.
In the language of the conformal metric coordinates (\ref{ansatzconf}) the $00$--component of
(\ref{ecualternativa}) reads
\begin{equation}
\label{ecuacionconforme00} f'' + 3f'^2=3H^2 -\frac{2}{3}\kappa_5^2\,\Lambda_{5} e^{2f}
-\frac{\kappa_5^2\,e^{2f}}{3} \frac{ 2+(\nabla T)^2 }{\sqrt{1 +(\nabla T)^2}}V(T),
\end{equation}
since
\begin{equation}
R_{00}=-3H^2 +f'' +3f'^2\label{R00}
\end{equation}
and
\begin{equation}
\hat{T}_{00}= - \frac{V}{3} e^{2f}\frac{2+ (\nabla T)^2}{\sqrt{1+(\nabla T)^2}}. \label{T00hat}
\end{equation}
Now let the metric be perturbed as follows
\begin{equation}
ds_{p}^2 = (g_{A B}+h_{A B}) dx^{A}dx^{B}= (g_{\mu \nu}+h_{\mu \nu}) dx^{\mu}dx^{\nu} + e^{2f(w)}
dw^2, \label{ansatzpert}
\end{equation}
where $g_{A B}=(g_{\mu \nu},e^{2f(w)})=e^{2f(w)} \overline{g}_{A B}=e^{2f(w)}( \overline{g}_{\mu
\nu},1)$, $\overline{g}_{\mu \nu} =\mbox{diag}[(-1,\delta_{i j}a^2)]$, $\mu,\nu=0,\ldots, 3$ and
$i,j=1,2,3$. Since we are studying the tensorial sector of the metric fluctuations, we can ignore
the $h_{M5}$ perturbations and set them to zero. From them, the $h_{m5}$ components belong to the
vector sector, while the $h_{55}$ mode has scalar nature and couples to the fluctuations of the
scalar tachyon field $T$.

In this approach we shall impose the transverse traceless condition on these metric fluctuations:
$\nabla^{\mu}h_{\mu \nu} = h_{\;\,\mu}^{\mu} = 0,$ where $\nabla$ is the covariant derivative
operator with respect to the metric $g_{A B}$.

After some algebraic work, the linearized Einstein equations for the transverse traceless
fluctuation modes adopt the following form
\begin{equation}\label{ecuaceinsteinlineal}
\delta R_{A B}=\frac{2}{3}\kappa_{5}^2\,\Lambda_{5} h_{A B} + \frac{\kappa_{5}^2\,V}{3}
\frac{2+(\nabla T)^2}{\sqrt{1+(\nabla T)^2}} h_{A B}.
\end{equation}
At this stage, it is extremely useful to define a new fluctuation variable $h_{\mu
\nu}=e^{2f}\overline{h}_{\mu \nu}$. By making use of the conformal transformation for the Ricci
tensor, in terms of the barred fluctuations we have
\begin{equation}\label{rvariacion}
\delta R_{\mu \nu}=\delta \overline{R}_{\mu \nu} -\frac{3}{2}f' \overline{h}_{\mu \nu}'
-\overline{h}_{\mu \nu}(f'' +3 f'^{2})
\end{equation}
where all the barred quantities are computed with respect to the metric $\overline{g}_{A B}$. On
the another hand, this quantity can also be computed as follows (see, for instance, \cite{'tHooft})
\begin{equation}
\delta \overline{R}_{\mu \nu} = \frac{1}{2}\left( -\overline{\Box} \overline{h}_{\mu \nu} +
\overline{\nabla}_{A}\overline{\nabla}_{\mu}\overline{h}_{\nu}^{A} +
\overline{\nabla}_{A}\overline{\nabla}_{\nu}\overline{h}_{\mu}^{A} -
\overline{\nabla}_{\mu}\overline{\nabla}_{\nu}\overline{h}_{A}^{A} \right),
\end{equation}
where $\overline{\Box} \overline{h}_{\mu \nu}= \overline{g}^{\alpha
\beta}\overline{\nabla}_{\alpha} \overline{\nabla}_{\beta}\overline{h}_{\mu \nu}
+\overline{g}^{55}\overline{\nabla}_{5} \overline{\nabla}_{5}\overline{h}_{\mu \nu}$.
In order to compute the involved quantities in this expression,  recall that the non--null
Christoffel symbols are
\begin{eqnarray}
\overline{\Gamma}^{0}_{i j}=a^2 H \delta_{i j},\qquad \qquad \overline{\Gamma}^{i}_{j
0}=H\delta^{i}_{j}.\nonumber
\end{eqnarray}
Since $\overline{\Gamma}^{5}_{A B}=\overline{\Gamma}^{B}_{5 A}=0$ and $\overline{h}_{5 A}=0$ it follows that
\begin{eqnarray}
\overline{\nabla}_{5}\overline{\nabla}_{5}\overline{h}_{\mu \nu}=h_{\mu \nu}^{''}, \qquad \qquad
\overline{g}^{\alpha \beta}\overline{\nabla}_{\alpha} \overline{\nabla}_{\beta}\overline{h}_{\mu
\nu}=\overline{\Box}_{4}\overline{h}_{\mu \nu},
\end{eqnarray}
where $\overline{\Box}_{4}$ is the d'Alembert operator on $dS_4$. While it is evident that $h =0$
implies $\overline{h} = 0$ from their conformal relationship, it is not straightforward to see
that $\overline{\nabla}^{\mu}\overline{h}_{\mu \nu} = 0$ follows from $\nabla^{\mu}h_{\mu \nu} =
0$. Hereon we show that this is indeed the case. The definition of the covariant derivative
implies that
$$
\nabla^{\mu}h_{\mu \nu} = \overline{g}^{\mu \alpha}\nabla_{\alpha}\overline{h}_{\mu \nu} =
\overline{g}^{\mu \alpha}\left(\partial_{\alpha}\overline{h}_{\mu \nu} - \Gamma^{E}_{\alpha
\mu}\overline{h}_{E \nu}-\Gamma^{E}_{\alpha \nu}\overline{h}_{E \mu}\right).
$$
Likewise, from the relationship between the Christoffel symbols of two conformal metrics, for the case here
considered it follows that
$$
\Gamma^{A}_{B C}=\overline{\Gamma}^{A}_{B C}+\left(\delta^{A}_{B}\overline{\nabla}_{C}f +
\delta^{A}_{C}\overline{\nabla}_{B}f - \overline{g}_{B C}\overline{\nabla}^{A}f\right).
$$
From both of these relations one concludes that $\Gamma^{\alpha}_{\beta
\gamma}=\overline{\Gamma}^{\alpha}_{\beta \gamma}$ and, therefore, that
$\nabla^{\mu}h_{\mu \nu} =
\overline{\nabla}^{\mu}\overline{h}_{\mu \nu}=0$.
Since in a curved spacetime the double covariant derivatives do not commute, for the the metric
fluctuations one has (see \cite{Carroll}, for instance)
\begin{eqnarray}
\overline{\nabla}_{C}\overline{\nabla}_{\mu} \overline{h}^{C}_{\nu}=
\overline{\nabla}_{\mu}\overline{\nabla}_{C} \overline{h}^{C}_{\nu}+\overline{R}_{E \mu}
\overline{h}^{E}_{\nu}-\overline{R}^{E}_{\nu C \mu}\overline{h}^{C}_{E}.
\end{eqnarray}
By making use of the transverse conditions $\overline{\nabla}^{\mu}\overline{h}_{\mu \nu}=0$ and
the fact that $\overline{\Gamma}^{5}_{A B}=\overline{\Gamma}^{B}_{5 A}=0$, it reads
\begin{eqnarray}
\overline{\nabla}_{C} \overline{h}^{C}_{\nu}=\overline{\nabla}_{5}
\overline{h}^{5}_{\nu}+\overline{\nabla}_{\alpha} \overline{h}^{\alpha}_{\nu}=0, \qquad \qquad
\overline{\nabla}_{\mu}\overline{\nabla}_{C} \overline{h}^{C}_{\nu}=0.
\end{eqnarray}
Since $h_{5 A}=0$ and $\overline{R}_{5 A B C }=\overline{R}_{5 A}=0$, one gets
\begin{eqnarray}
\overline{\nabla}_{C}\overline{\nabla}_{\mu} \overline{h}^{C}_{\nu}= \overline{R}_{\alpha \mu}
\overline{h}^{\alpha}_{\nu}-\overline{R}^{\alpha} _{\nu \beta \mu}\overline{h}^{\beta}_{\alpha}.
\end{eqnarray}
Moreover, for a $dS_4$ spacetime the following relations hold
\begin{eqnarray}
\overline{R}_{\mu \nu \alpha\beta } = H^2 (\overline{g}_{ \mu \alpha}\overline{g}_{\nu \beta} -
\overline{g}_{\mu \beta}\overline{g}_{\nu \alpha}),\qquad \qquad
\overline{R}_{\mu \nu}=3H^2
g_{\mu \nu}.
\end{eqnarray}
Hence, we get the following result for the linear variation of the Ricci tensor
\begin{equation}
\delta \overline{R}_{\mu \nu} = \frac{1}{2}\left( -\overline{\Box}_{4} \overline{h}_{\mu \nu} -
\overline{h}_{\mu \nu}^{''} +8H^{2}\overline{h}_{\mu \nu} \right).
\end{equation}
Finally, by making use of the relations (\ref{ecuacionconforme00}), (\ref{ecuaceinsteinlineal}),
(\ref{rvariacion}) and the last expression we have
\begin{eqnarray}
\overline{\Box}_{4}\overline{h}_{\mu \nu}+\overline{h}_{\mu \nu}'' + 3f' \overline{h}_{\mu \nu}' -
2H^{2}\overline{h}_{\mu \nu}=0, \label{flucteqn}
\end{eqnarray}
under the imposed transverse and traceless conditions $\overline{\nabla}^{\mu}\overline{h}_{\mu
\nu}=0, \ \overline{h}^{\alpha}_{\alpha}=0$.

Furthermore, by performing the following separation of variables for the metric fluctuations $\bar h_{\mu
\nu}=e^{-\frac{3}{2}f(w)}\Psi(w)\phi_{\mu\nu}(x)$, it leads (\ref{flucteqn}) into a
Schr\"odinger--like equation along the extra dimension:
\begin{equation}
\left(-\partial^{2}_{z}+V_{QM}-m^{2} \right)\Psi(w)=0,
\label{Scheqn}
\end{equation}
where the analogue quantum mechanical potential $V_{QM}$ reads
\begin{equation}
V_{QM}=\frac{9}{4}f^{'2}+\frac{3}{2}f^{''}. \label{VQM}
\end{equation}
Now, the 4D equation indited  from (\ref{flucteqn}) is
\begin{equation}
\left(-\partial^{2}_{t}-3H\partial_{t}+e^{-2Ht}\nabla^2-2H^2 \right)\phi(x)=-m^{2}\phi(x),
\end{equation}
where $m^{2}$ represents the mass that a 4D observer perceives in a de Sitter spacetime
\cite{dsmass, wang}. The indices of the function $\phi_{\mu\nu}(w)$ were omitted, for
convenience.

By substituting the expression for the warp factor (\ref{fw}) into (\ref{VQM}) the analog quantum
mechanical potential adopts the form of a modified P\"oschl--Teller potential that reads
\begin{equation}
V_{QM}=\frac{3H^{2}}{4}\left[3-7{\rm sech}^{2}(2Hw)\right]. \label{VQMw}
\end{equation}
The fact that this potential possesses a definite positive asymptotic value ensures the existence
of a mass gap in the graviton spectrum of KK massive fluctuations determined by
$\frac{9 H^{2}}{4},$ or equivalently $m=\frac{3H}{2}$ \cite{wang, PS, bhnqrs, corrNL3, cuco, GLWF}.

By performing the following rescaling of the fifth coordinate $v=2Hw$ we recast (\ref{Scheqn})
into
\begin{equation}
 \left(-\partial^{2}_{v} - \frac{21}{16}{\rm sech}^{2}v \right)\Psi(v) =
 \left(\frac{m^{2}}{4H^{2}} - \frac{9}{16}\right)\Psi(v),
 \label{Scheqnv}
\end{equation}
which can be directly compared to the canonical form of the classical eigenvalue problem for the
Schr\"odinger equation with a modified P\"oschl--Teller potential
\begin{equation}
 \left[-\partial^{2}_{v} - n(n+1){\rm sech}^{2}v \right]\Psi(v) =
 E\,\Psi(v)
 \label{Schcan}
\end{equation}
with $n=3/4$ and $E=\frac{m^{2}}{4H^{2}} - \frac{9}{16}$. Since $n<1$ there is just one bound
state in the mass spectrum of KK metric fluctuations: the zero mode massless state which accounts
for the massless 4D graviton. To the best of our knowledge, this is the first model that presents
just a {\it single} bound state in this kind of spectra within the framework of thick braneworlds.
Usually one encounters a second bound state which represents a massive KK excitation with no clear
physical interpretation (see \cite{gremm}, \cite{wang, PS, bhnqrs, corrNL3, cuco, GLWF}, for
instance). Eq.(\ref{Scheqnv}) can be integrated for an arbitrary mass and possess the following
general solution:
\begin{equation}
 \Psi(w)=C_1\,P^{\mu}_{3/4}\left(\tanh(2Hw)\right)+
 C_2\,Q^{\mu}_{3/4}\left(\tanh(2Hw)\right)
\end{equation}
where $C_1$ and $C_2$ are constants, while $P^{\mu}_{3/4}$ and $Q^{\mu}_{3/4}$ are associated
Legendre functions of first and second kind, respectively, with degree $\nu=3/4$ and order
$\mu=\sqrt{\frac{9}{16}-\frac{m^{2}}{4H^{2}}}$. For the massless case the zero mode has
$\mu=\nu=3/4$ and the exact solution can be expressed as
\begin{equation}
 \Psi_{0}(w)=-k_1\,\left[P^{3/4}_{3/4}\left(\tanh(2Hw)\right)+
 \frac{2}{\pi}\,Q^{3/4}_{3/4}\left(\tanh(2Hw)\right)\right],
 \label{zeromode}
\end{equation}
where now $k_1=-C_1>0$ and we have set $C_2=2C_1/\pi$ in order to get a localized configuration.
This bound state is physically interpreted as a stable graviton localized on the brane, since
there are no states with negative squared masses due to the positive definite character of the
zero mode (\ref{zeromode}) and the structure of the potential (\ref{Schcan}). The behavior of
both the modified P\"oschl--Teller potential and the graviton zero mode is displayed in Fig. 3.

\begin{figure}[htb]
\begin{center}
\includegraphics[width=7cm]{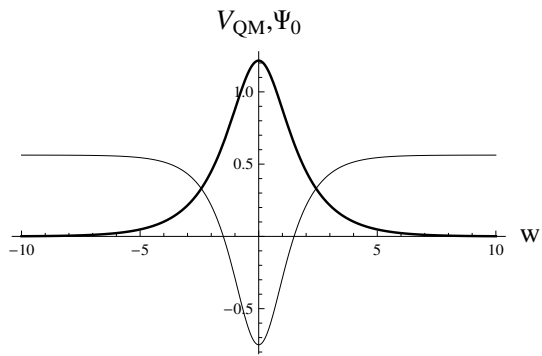}
\end{center}\vskip -5mm
\caption{The profile of the modified P\"oschl--Teller potential (thin line) and the localized 4D
graviton zero mode (thick line) along the fifth dimension. Here we have set $c=0$, $H=1/2$ and
$k_1=1$ for simplicity.} \label{fig_zeromodeinVqm}
\end{figure}

Finally, there is also a continuum of KK massive modes in the spectrum that starts from $m\geq
3H/2$ and is described by eigenfunctions with imaginary order $\pm\mu=\pm i\rho$ \cite{bhnqrs}:
\begin{equation}
\Psi_{m}(w)= \sum_{\pm}C_{\pm}\,P^{\pm i\rho}_{3/4}\left(\tanh(2Hw) \right), \label{massmod}
\end{equation}
where $C_{\pm}$ are arbitrary constants and $\rho=\sqrt{\frac{m^{2}}{4H^{2}}-\frac{9}{16}}.$ These
KK massive modes must behave as plane waves asymptotically \cite{wang,PS}. This fact can be seen
by considering masses $2m > 3H$ and taking into account that the constants $C_{\pm}$ depend on
$\rho$ in general, i.e.
\begin{eqnarray}
\Psi^{\mu}_{\pm}(w)= C_{\pm}(\rho)P^{\pm i\rho}_{3/4} \left(\tanh(2Hw) \right). \label{Psipm}
\end{eqnarray}
We further make an expansion for large $w$ of the argument of the associated Legendre functions of
first kind:
\begin{equation}
\tanh(2Hw) \approx 1 - 2e^{-4Hw} \label{asymptanh}
\end{equation}
and compute the asymptotic behavior of $P^{\pm i\rho}_{3/4}\left(\tanh(2Hw)\right)$ according to
 Eq.(8) of Sec. (3.9.2) in \cite{bateman}:
\begin{eqnarray}
P^{\pm i\rho}_{3/4}\left(\tanh(2Hw)\right) \sim \frac{1}{\Gamma(1\mp i\rho)}e^{\pm 2iH\rho w}.
\label{Pasymp}
\end{eqnarray}
The normalization condition of (\ref{Psipm}) in the plane wave sense leads to the following
normalization constants
\begin{eqnarray}
C_+(\rho)=C_-(\rho) =\frac{\vert\Gamma(1+i\rho)\vert}{\sqrt{2\pi}} \label{Cpmfin}
\end{eqnarray}
since $\vert{\Gamma(1-i\rho)}\vert=\vert{\Gamma(1+i\rho)}\vert$. Thus, by substituting
(\ref{Pasymp}) and (\ref{Cpmfin}) into (\ref{Psipm}) we obtain the asymptotic behavior of these
associated Legendre functions:
\begin{equation}
\Psi_{\pm}^{\mu}(w)\sim \frac{1}{\sqrt{2\pi}}e^{\pm i2H\rho w}
\end{equation}
which corresponds to plane waves as expected.

Once  an analytical expression for the KK massive modes was obtained, we should be able to compute
the corresponding small corrections to Newton's law due to these 5D massive modes. This is
achieved by taking the thin brane limit $H\rightarrow\infty$, locating a probe mass, $M_1$, in
the center of the brane in the transverse direction. Subsequently, by computing the gravitational
potential generated by this particle, felt by another massive particle with mass $M_2$. The
corrections to the Newtonian potential generated by massive gravitons in the thin brane limit can
be expressed as follows \cite{csakietal}
\begin{eqnarray}
U(r)\sim \frac{M_{1}M_{2}}{r}\left(G_{4}+M_{*}^{-3}\int_{m_{0}}^{\infty}dme^{-mr}
\left|\Psi^{\mu(m)}(w_{0})\right|^{2}\right) = \frac{M_{1}M_{2}}{r}\left(G_{4}+\triangle
G_{4}\right), \label{U}
\end{eqnarray}
where $w = w_{0}$ sets the position where the brane is located, $m_{0}=3H/2$ for our case, $G_{4}$
is the gravitational 4D coupling constant and $\Psi^{\mu}(w_{0})$ denotes the continuum of KK
massive modes that must be integrated over their masses in order to get the searched corrections.

Now we proceed to calculate $\left|\Psi^{\mu}(0)\right|^2$ at $w_0=0$ and get
\begin{eqnarray}
\left|\Psi^{\mu}(0)\right|^2 = \left|\frac{\Gamma\Bigl(1+i\rho\Bigr)}
{\Gamma\Bigl(\!\frac{11}{8}\!+\!\frac{i\rho}{2}\Bigr)
\Gamma\Bigl(\frac{1}{8}\!+\!\frac{i\rho}{2}\Bigr)}\right|^2\!, \label{modulesquared}
\end{eqnarray}
where we set $\nu=3/4$ and made use of the following relation for the associated Legendre
functions of first kind \cite{GR}:
\begin{eqnarray}
P^{\mu}_{\nu}(0)=\frac{2^{\mu}\sqrt{\pi}}{\Gamma\Bigl(\frac{1-\nu-\mu}{2}\Bigr)
\Gamma\Bigl(1+\frac{\nu-\mu}{2}\Bigr)}. \label{P0}
\end{eqnarray}
Substitution of (\ref{modulesquared}) into the second term of (\ref{U}) yields the following
expression for $\Delta G_4:$
\begin{eqnarray}
\Delta G_4\!=M_{\ast}^{-3}\!\int_{m_0}^{\infty}\!dm\,e^{-mr}
\left|\frac{\Gamma\Bigl(1+i\rho\Bigr)} {\Gamma\Bigl(\!\frac{11}{8}\!+\!\frac{i\rho}{2}\Bigr)
\Gamma\Bigl(\frac{1}{8}\!+\!\frac{i\rho}{2}\Bigr)}\right|^2\!. \label{DG4}
\end{eqnarray}
In order to calculate this integral it is convenient to perform a change of variable from $m$ to
$\rho$:
\begin{eqnarray}
\Delta G_4 = 2M_{\ast}^{-3}H
\int_{0}^{\infty}\frac{e^{-2\sqrt{\rho^2+\frac{9}{16}}Hr}}{\sqrt{1+\frac{9}{16\rho^2}}}
\left|\frac{\Gamma\Bigl(1+i\rho\Bigr)} {\Gamma\Bigl(\!\frac{11}{8}\!+\!\frac{i\rho}{2}\Bigr)
\Gamma\Bigl(\frac{1}{8}\!+\!\frac{i\rho}{2}\Bigr)}\right|^2\,d\rho. \label{DG4fin}
\end{eqnarray}
This integral can be calculated in the thin brane limit $H\longrightarrow\infty$, where it is
dominated by the region of small -- $\rho$. Thus (\ref{DG4fin}) can be well--approximated by
expanding the factor that multiplies the exponential at $\rho = 0$ \cite{bhnqrs}. Thus, the
contribution to Newton's law made by the continuum of KK massive modes in the thin brane limit
reads:
\begin{equation}
\triangle G_{4}\sim \frac{M_{*}^{-3}}{\left| \Gamma(\frac{11}{8})\Gamma(\frac{1}{8})
\right|^2}\frac{e^{-\frac{3}{2}Hr}}{r}\left(1+O\left(\frac{1}{Hr}\right)\right).
\end{equation}
These corrections are exponentially suppressed as in other braneworld models with an induced 4D
Minkowski \cite{corrNL1}, (see also \cite{bhnqrs}--\cite{corrNL3}) or de Sitter metric
\cite{cuco,GLWF}. This result shows that the form of the corrections to Newton's law coming from
the extra dimension is quite robust against the explicit form of the KK massive modes as well as
the type of the field that couples to gravity to generate the braneworld model.

\section{Discussion}

In this paper we have presented a thick braneworld model generated by a tachyon scalar field
coupled to gravity with a bulk cosmological constant and a de Sitter metric induced on the brane.
We were able to obtain an exact solution with a decaying warp factor that gives rise to
localization of 4D gravity when studying the metric fluctuations. The structure of the
corresponding graviton spectrum is novel in the sense that it contains just one bound state (the
massless zero mode) which is physically interpreted as the 4D graviton, separated by a mass gap
from a continuum of KK massive excitations. We provided an explicit expression for these KK
massive modes, a fact that enables us to analytically compute the corrections to Newton's law
coming from the extra dimension. As we mentioned above, these corrections are exponentially
suppressed and coincide (up to a coefficient factor) with previous results reported in the
literature within the framework of similar braneworld models with the same 4D de Sitter induced
metric. This result reflects the robustness of the form of the corrections to Newton's law since
they are computed for KK massive modes with different form compared to other braneworld models
previously reported.

When analyzing the curvature scalar of our model we realize that it is positive definite on the
whole patch and asymptotically vanishes even when we have a negative cosmological constant. Thus,
our de Sitter tachyonic thick braneworld model interpolates between two 5D Minkowski spaces. On
the other hand, we should remember that our analysis is carried out in a local basis. Thus, the
study of the global structure of our 5D spacetime, which involves the continuation of our
coordinate system defined by (\ref{ansatz}), could lead to a more complex picture in which the
curvature is positive in our chart, but is asymptotically negative, for instance. The study of
this interesting properties is beyond the scope of the present paper and will be reported
elsewhere.

Since here a braneworld with an induced 4D de Sitter metric is considered, in principle, we are
able to reproduce the early inflation and accelerated expansion epochs of our universe within our
model. However, a more general and realistic ansatz for our setup that attempts to describe the
late time behavior of our 4D universe should involve both a time--depending tachyon field and a
time--depending warp factor, since from the cosmological viewpoint one needs to obtain scale
factors that reproduce in a better way (closer to the observations) the accelerated expansion of
the universe (which takes into account its dark matter component) than the de Sitter metric does.
This line of research is under current investigation and could lead to interesting new results in
cosmology.

Another interesting issue that should be approached is related to the stability of this braneworld
configuration. This study involves computing the field equations of the perturbed tachyonic scalar
field coupled to the scalar modes of the metric fluctuations in the linear approximation. Recall
that the scalar sector of the perturbed metric must be defined with respect to the transformations
associated to the $dS_4$ symmetry group. This is a rather involved task that is already under
performance and will be reported in the near future. In the lack of such a rigorous analysis we
can instead study the stability of the brane in the limit of small gradient for the tachyonic
scalar field. This is done by considering the dynamics of scalar perturbations when taking into
account the back reaction of the brane itself.

Thus, we shall consider a sort of ``slow--roll" approximation in the action (2.1). In this limit,
the kinetic term of the tachyon field is considered small compared to the self--interaction
potential. However, we shall instead assume a small gradient approximation $(\nabla T)^2<<1$, so
that the action for the tachyon field could be expanded up to quadratic first derivative terms
\cite{zwie}:
\begin{equation}
S=-\int d^{5}x V(T)\left(1+\frac{1}{2}g^{AB}\partial_A T\partial_B T+...\right).
\end{equation}
This action can be further written as an action for a standard scalar field:
\begin{equation}
S=\int d^{5}x\left(-\frac{1}{2}g^{AB}\partial_A\varphi\partial_B\varphi-V(\varphi)+...\right),
\end{equation}
where the new scalar field $\varphi=\varphi(T)$ and we have assumed that
\begin{equation}
V=\left(\varphi_T\right)^2=\left(\frac{\partial\varphi}{\partial T}\right)^2 \label{VTphi}
\end{equation}
in order to identify $\varphi_T\partial_A T$ with $\partial_A\varphi$. Thus, in the small gradient
approximation, the tachyon dynamics is described by the action for the standard scalar field
$\varphi$. Moreover, from the latter relation (\ref{VTphi}) one can easily read the expression for
its first derivative with respect to the extra coordinate in terms of the tachyonic field and its
first derivative: $\varphi'=\sqrt{V}\,T'.$

In order to study the stability behaviour of our tachyonic brane in the small gradient
approximation, and under the ansatz (2.11), we must consider the regime $(\nabla T)^2 <<1$.

The analysis of scalar perturbations $\varphi=\varphi_0 + \delta\varphi$ for fluctuated
non--singular solutions of the Einstein equations with non--trivial standard scalar fields and
self--interaction potentials was performed in \cite{KKS}. These authors computed the effective
potentials for the master Schr\"odinger--like equation of scalar perturbations for a family of
smooth solutions corresponding to different bulk curvatures and found that they are positive
definite as well as the spectrum of massive modes, consequently, these systems were found to be
stable under small perturbations.

We can perform a similar analysis in the small gradient approximation of our relevant tachyonic
thick braneworld. Thus, by quoting the relevant Schr\"odinger--like equation obtained in
\cite{KKS} for scalar perturbations $F(w,x^{\mu})$ (see this work for details):
\begin{equation}
-F''(w,x^{\mu})+ V_{eff}(w)\,F(w,x^{\mu}) = m^{2}F(w,x^{\mu}), \label{scalarScheqn}
\end{equation}
where we have taken into account that $\delta\varphi=\delta\varphi(F,F')$, $m$ is the mass of the
scalar perturbation in a 4D de Sitter space \cite{dsmass} and the effective potential $V_{eff}$
reads
\begin{equation}
V_{eff} = - \frac{5}{2}f'' + \frac{9}{4}{f'}^2 +f'\frac{\varphi_0''}{\varphi_0'}
-\frac{\varphi_0'''}{\varphi_0'} +2\left(\frac{\varphi_0''}{\varphi_0'}\right)^2 -4H^2,
\label{Veff}
\end{equation}
where $\varphi_0'=\sqrt{V}\,T'.$

The form of this effective potential corresponds to a potential barrier which asymptotically
approaches the positive value $\frac{17H^2}{4}$ and possesses a maximum at $\frac{51H^2}{10}$ (see
Fig. 4). Thus, the spectrum of massive scalar fluctuations is positive definite, a fact that
indicates that we have a stable scalar field configuration around the origin, where the 3--brane
is located, in the small gradient approximation for the scalar field.

However, it should be pointed out that since $(\nabla T)^2 = e^{-2f}(T')^2 = \frac{1}{4}\mbox{sech}^2(Hw)$ (here we have set $c=0$), at the
origin we have a $25\%$ of a unity instead of a small quantity. This means that our result seems
to have sense in the small gradient approximation unless the potential barrier configuration turns
into a potential well sufficiently deep in order to get unstable modes in the massive spectrum of
scalar perturbations when one computes the exact dynamical equation and, hence, is not so robust.

\begin{figure}[htb]
\begin{center}
\includegraphics[width=7cm]{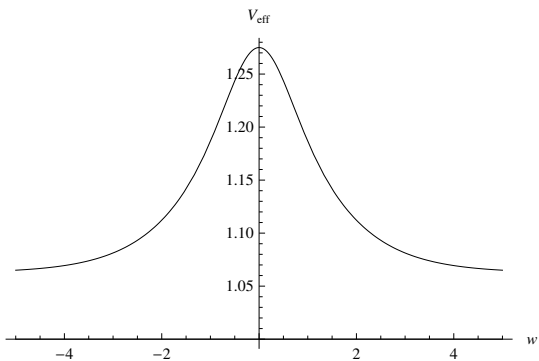}
\end{center}\vskip -5mm
\caption{The shape of the approximated self--interaction potential for the tachyonic scalar field
in the small gradient approximation. We set here $c=0$, $H=1/2$ and $2\kappa_{5}^2=1$ for
simplicity.} \label{fig_approxptl}
\end{figure}

\section*{Acknowledgments}
GG and AHA thank SNI for support. AHA is grateful to U. Nucamendi for useful discussions and to
the staff of the ICF, UNAM for hospitality; the research of AHA, DMM and RRML was supported by
grant CONACYT 60060-J. DMM acknowledges a PhD grant from UMSNH and a DGAPA-UNAM postdoctoral fellowship, while RRML acknowledges a PhD grant from CONACYT. GG, AHA, DMM and RRML gratefully acknowledge support from ``Programa de Apoyo a Proyectos de Investigaci\'on e Innovaci\'on Tecnol\'ogica” (PAPIIT) UNAM, IN103413-3, {\it Teor\'\i as de Kaluza-Klein, inflaci\'on y perturbaciones gravitacionales}.
RR is grateful to E. Capelas de Oliveira for fruitful discussions and to Conselho Nacional de
Desenvolvimento Cient\'{\i}fico e Tecnol\'ogico (CNPq) grants 476580/2010-2 and 304862/2009-6 for
financial support.

\end{document}